
\magnification=\magstep1
\baselineskip=14.0pt
\vsize=9.0 true in

\def\bxm{X^m (f_j)}
\def\bxi{X^I (f_j)}
\def\bxip{{X^I}' (f_j)}
\def\lxm#1{{m}_{#1}}
\def\lxi#1{{n}_{#1}}
\def\Dt{\Delta t}
\def\smk{\sum_{k=0}^{N-1}}


\def\til{\mathaccent"0365}

\def\ie{{\it i.e. }}
\def\cf{{\it cf. }}

\def\aproxgt{\mathrel{%
	\rlap{\raise 0.511ex \hbox{$>$}}{\lower 0.511ex \hbox{$\sim$}}}}
\def\aproxlt{\mathrel{%
	\rlap{\raise 0.511ex \hbox{$<$}}{\lower 0.511ex \hbox{$\sim$}}}}


\def\apj#1#2#3#4{
\noindent\hangindent=20pt\hangafter=1{
#1~ #2, {ApJ}, {#3}, #4}}
\def\apjsup#1#2#3#4{
\noindent\hangindent=20pt\hangafter=1{
#1~ #2, {ApJS}, {#3}, #4}}


\def\book#1#2#3#4#5#6{
\noindent\hangindent=20pt\hangafter=1{
#1~ #2, {#3}, (#4:#5), p. #6}}

\def\bookpl#1#2#3#4#5{
\noindent\hangindent=20pt\hangafter=1{
#1~ #2, {#3}, (#4:#5)}}


\centerline{\bf CAN WE IDENTIFY LENSED GAMMA RAY BURSTS?}
\bigskip
\centerline{\sl Michael A. Nowak and Scott A. Grossman}
\smallskip
\centerline{CITA, 60 St. George St., Toronto, Ontario ~M5S 1A7}

\baselineskip=12.0pt

\bigskip
\centerline{ABSTRACT}
\medskip

A gravitationally lensed gamma-ray burst (GRB) would appear as
multiple bursts with identical light curves, separated in time and
differing only by the scaling of their amplitudes.  In reality, the
detected bursts will be noisy, and therefore they may be
difficult to identify as lensed images.  Furthermore, faint,
intrinsically similar, yet distinct light curves may be falsely
identified as lensing events. In this paper we develop two statistical
tests to distinguish noisy burst light curves.  We use Fourier
analysis techniques to analyze the signals for both intrinsic
variability and variability due to noise.  We are able to determine
the noise level, and we compare the bursts only at frequency channels
that are signal dominated.  Utilizing these methods, we are able to
make quantitative statements about whether two bursts are distinct.
We apply these statistics to scaled versions of two subbursts of GRB
910503--- subbursts previously investigated by Wambsganss (1993) using
a different statistical test. We find that  our methods are able to
distinguish these bursts at slightly smaller amplitudes than those at
which Wambsganss's method succeeds. We then apply our techniques to
``candidate" lensing events taken from the BATSE catalogue, and we find
that nearly all of them, except for the very shortest ones (durations
$\aproxlt 0.3 ~s$), are distinguishable.  We therefore expect that a
majority of bursts will be disinguishable from one another.

\baselineskip=14.0pt

\bigskip
\centerline{1. Introduction}
\medskip

At the Second Huntsville Gamma Ray Burst Workshop, the question of
whether gamma ray bursts (GRBs) are galactic or cosmological in
origin was put to a vote, and the attendants were split 60\%-40\% in
favor of the cosmological scenario (\cf Wang 1994).  More than 20
years after their discovery, and after more than a hundred
proposed models (\cf Nemiroff 1994a), not even the location of the
bursts are agreed upon! There are strong arguments to
support both views (\cf Blaes 1994 and Hartmann 1994
for reviews of the observations and theories).  However, simple
standard-candle cosmological scenarios, wherein GRBs have a constant
burst rate per comoving volume per comoving
time, appear to fit the data quite well ({\it e.g.} Fenimore et al.
1993).  These models are attractive because they naturally explain the
observed isotropy and flux distribution of the bursts.  These models,
however, only show consistency with the data, and do not provide
proof of the cosmological scenario.

As first suggested by Paczy\'nski (1986), perhaps the only unambiguous
verification of a cosmological origin for GRBs would be the detection
of a gravitationally lensed burst.  The multiple images of a lensed
GRB would have identical light curves (to within the
noise and an overall magnification) that arrive from the
same position on the sky; however, the individual bursts would be
separated in time. A number of authors have investigated the
possibility of GRB lensing. Mao (1992) considered lensing by galaxies
and calculated the expected time delays between lensed bursts.
He concluded that the Burst and Transient Source Experiment (BATSE)
aboard the Compton Gamma Ray Observatory (CGRO) has a significant
chance of observing a lensing event during its lifetime.  More
recently, Grossman \& Nowak (1994) have refined the calculation of
Mao (1992) and they find a substantially smaller probability that
BATSE will observe GRBs lensed by galaxies (although a suitably
designed instrument likely could detect a lensed burst).  Blaes \&
Webster (1992) considered lensing by point masses, Narayan \&
Wallington (1992) investigated the possibility of determining lens
parameters from an observed event, and Nemiroff et al. (1993a,b)
have used the absence of any observed lensing events to put rather
loose constraints upon the distances to bursts and upon the density of
$10^6 ~ M_\odot - 10^8 ~ M_\odot$ objects in the universe.  [Other
authors--- namely Paczy\'nski (1987), Gould (1992), and Mao (1993)---
have considered micro- and ``femto-" lensing.  These events may be
manifest within a single light curve in some cases. In this work,
however, we only consider the possibility of identifying lensing events
in distinct GRB light curves.]

The discussion above presumes that we will actually
{\it recognize} a lensing event when we see it.  As first pointed
out by Wambsganss (1993), such recognition may be very difficult.  The
BATSE light curves are comprised of an intrinsic source signal, a
background, and noise.  Only the source component will be identical
(to within an overall scale factor) among the lensed images.  It is
possible that the background and noise may overwhelm our ability to
identify a lensing event.  Another potential problem is that two
distinct but intrinsically similar light curves, if faint enough, may
be falsely identified as a lensing event.  To address these issues,
Wambsganss (1993) developed a test to determine whether or not two
gamma ray bursts with similar time profiles could be statistically
distinguished from each other. His method essentially compared the
variance of the difference of the two light curves (where one light
curve is scaled and time-shifted compared to the other) with the
variance due to Poisson noise.  If the variance was  consistent with
that due to Poisson noise only, the bursts were said to be
statistically identical.  Wambsganss then applied his test to two
intrinsically similar but distinct BATSE light curves and found that
as the amplitudes of the bursts were reduced (keeping the backgrounds
constant, and adding an appropriate amount of counting noise), the
light curves became indistinguishable.  Furthermore, the light curves
became indistinguishable at a level well above the BATSE detection
threshold. Wambsganss (1993) thus concluded that it
will be very difficult in practice to identify lensed GRBs.

Wambsganss's method, however, used every time bin of the light
curves, whether the bins were noise dominated
or not.  No attempts were made to minimize the noise.
Furthermore, Wambsganss's criteria for burst distinction were more
subjective than quantitative. (No significance values for the
comparisons are presented in Wambsganss [1993].) In this paper, we use
Fourier analysis techniques to analyze bursts for both intrinsic
variability and variability due to noise.  We are able to determine
the noise level in the Fourier domain, and compare burst light
curves only at frequency channels that are signal dominated.  We thus
do not dilute the signal with excessively noisy data. We quantify the
confidence with which we distinguish two light curves.  In
\S 2 we use the properties of Fourier transforms to define the two
statistics that we use to compare light curves.  In \S 3 we apply our
statistics to the same scaled light curves that Wambsganss (1993)
considered. Utilizing our methods, we can distinguish these two
bursts at slightly smaller amplitudes than those at which Wambsganss's
method succeeded.  We also discuss how the scaled amplitudes compare
to the BATSE detection threshold.  In \S 4 we apply our techniques to
``candidate" lensing events taken from the BATSE burst catalogue.  We
find that nearly all of these bursts, except for the very shortest
ones (durations $\aproxlt 0.3 ~s$), are easily distinguishable.  In \S
5 we summarize our results and discuss their implications for lensing
calculations.

\bigskip
\centerline{2. Comparison of Noisy Light Curves}
\medskip

Assume that we have two light curves, $s_g(t)$ and $s_h(t)$, comprised
respectively of intrinsic signals, $g(t)$ and $h(t)$, and
noise, $n_g(t)$ and $n_h(t)$.  We assume that the noise is counting
noise, and therefore it is uncorrelated among time bins.  We have
$$\eqalign{s_g(t) &= g(t) + n_g(t) ~~, \cr
           s_h(t) &= h(t) + n_h(t) ~~.} \eqno(2.1)$$
We shall assume that the above signals have the (assumed known)
background subtracted. If the intrinsic signals are scaled copies of
each other such that $g(t) = A~ h(t+\tau_d)$, where $A$ and $\tau_d$
are constants, then the normalized correlation function, defined by
$$ c(\tau) \equiv {{\int^\infty_{-\infty} g(t) h(t-\tau) ~dt}\over{
     {\sqrt{\int^\infty_{-\infty} g(t) g(t) ~dt}}
     {\sqrt{\int^\infty_{-\infty} h(t) h(t) ~dt}}}} ~~,
     \eqno(2.2)$$
is equal to unity if $\tau =  - \tau_d$.  The measured signals $s_g(t)$
and $s_h(t)$ include uncorrelated noise, and the resulting
correlation will be less than unity. The question we need to address
is: how much of a deviation from unity is significant, indicating that
$g(t)$ and $h(t)$ are not copies of each other?

Analysis of the correlation function is simplest
in the Fourier frequency domain.  Using the Correlation
Theorem  (\cf Davenport \& Root 1978), we define the cross power
spectral density (CPD),
$$ C(f) \equiv
     \int^\infty_{-\infty} \left ( \int^\infty_{-\infty} g(t) h(t-\tau)
     ~dt \right ) ~
     e^{i 2 \pi f \tau} ~d\tau  ~~, \eqno(2.3)$$
where $G(f)$ and $H(f)$ are the complex Fourier transforms of $g(t)$
and $h(t)$, respectively.  (Here and
throughout we shall use capital letters to denote functions in the
frequency domain, and lower case letters to denote functions in the
time domain.  We will suppress the functional dependences on $f$ and
$t$ whenever there is little chance of confusion.)  If $g(t) = A~
h(t+\tau_d)$, then
$$ C(f) = A | H(f) |^2 e^{- i 2 \pi f \tau_d } ~~. \eqno(2.4)$$
Note that the CPD time delay, defined as the complex phase of the CPD
divided by $2 \pi f$, is constant. In addition, the power spectral
densities (PSDs, \ie squared Fourier amplitudes) of the individual
signals are proportional to each other. In analogy to the normalized
correlation function, we find it convenient to use a Fourier transform
normalization such that the amplitude of the transform is independent
of the amplitude of the signal in the time domain (\cf Appendix A).
Therefore, in the absence of noise, lensed bursts will have identical
PSDs.

In reality, the measured
signals will be noisy, and therefore the PSDs will differ from
one another and the CPD time delay will not be constant. The advantage
of working in the frequency domain is that we can estimate the level of
the noise (which we do in Appendix A), and determine which frequencies
are above this noise level.  The criteria for lensed signals,
that is consistent Fourier amplitudes and coherent phases, must hold
for the signal dominated frequencies.  Below we derive error
statistics that measure the statistical significance of deviations
from perfect correlation.

In the following sections we shall construct ``optimized"
Fourier transforms from the measured transforms, and then use these
optimized transforms in our statistical tests.  From our
knowledge of the statistical properties of the noise (\cf Appendix
A) and our measurements of the signal,  we can construct ``optimal
filters" that when multiplied by our measured transforms yield
estimates of the ``true" transforms.  These optimal filters
statistically minimize the difference between the true and estimated
transforms. The literature and history of optimal filtering is vast
(\cf Whalen 1971); however, nothing presented in this work goes beyond
the scope of the discussion found in Press et al. (1993).  Below we
consider separately applications of optimal filtering to the PSDs and
CPDs of a pair of light curves.

\smallskip
\centerline{\it a) Comparison via Power Spectral Densities}
\smallskip

The Fourier transforms of equation (2.1) are given by
$$\eqalign{S_g(f) &= G(f) + N_g(f) \cr
           S_h(f) &= H(f) + N_h(f)  ~~. } \eqno(2.5)$$
In the following analysis, we adopt a Fourier
transform normalization in a light curve, based upon the integrated
counts, such that PSD amplitudes are  independent of the intensity of
the burst.  Thus,  for lensed bursts we expect $G^2 - H^2 =0$. We
measure only $S_g^2$ and $S_h^2$ however, and obtain only
statistical estimates of $N_g^2$ and $N_h^2$ (\cf Appendix A).  From
our measurements we wish to construct ``best estimates" of the
intrinsic PSDs, which we call $\til {G}^2$ and $\til {H}^2$.
These estimates are related to the measured signals according to
$$\eqalign{\til {G}^2 &\equiv S_g^2 ~\Phi_g   \cr
           \til {H}^2 &\equiv S_h^2 ~\Phi_h  ~~, } \eqno(2.6)$$
where the optimal filters, $\Phi_g$ and
$\Phi_h$, are functions that minimize $\langle ( \til {G}^2 - G^2 )^2
\rangle$ and $\langle ( \til {H}^2 - H^2 )^2 \rangle$.  The
brackets denote an ensemble average.
Expanding these functions we find
$$\langle ( \til {G}^2 - G^2 )^2 \rangle ~=~  \left [ (G^2 +
     \langle N_g^2 \rangle )^2 + 2 G^2 \langle N_g^2 \rangle \right ]
     ~\Phi_g^2 ~-~ 2 G^2 (G^2 + \langle N_g^2 \rangle ) ~\Phi_g ~+~
     G^4 ~~, \eqno(2.7)$$
where we have used the fact that $\langle N_g \rangle =0$.
Equation (2.7) is minimized at each frequency if
$$\Phi_g = {{G^2 (G^2 + \langle N_g^2 \rangle)} \over
     {(G^2 + \langle N_g^2 \rangle)^2 + 2 G^2 \langle N_g^2 \rangle}}
     \approx {{G^2 S_g^2}\over{S_g^4 + 2 G^2 \langle N_g^2 \rangle}}
     ~~. \eqno(2.8)$$
The approximation above uses $S_g^2 \approx G^2 + \langle N_g^2
\rangle$.
We obtain a similar expression for $\Phi_h$.  Note that $\Phi_g
\sim 1$ in the signal dominated regime, where $S_g \approx G \gg N_g$,
and $\Phi_g \sim G^2/\langle N_g^2 \rangle$ in the noise dominated
regime, where $S_g \approx N_g \gg G$. In practice, we calculate
the filters by taking $G^2 \approx S_g^2 - \langle N_g^2 \rangle$ and
$H^2 \approx S_h^2 - \langle N_h^2 \rangle$. Substituting the
optimal filter (2.8) into equation (2.7), we obtain
$$\langle ( \til {G}^2 - G^2 )^2 \rangle ~=~
     {{2 ~G^6 \langle N_g^2
     \rangle} \over {(G^2 + \langle N_g^2 \rangle )^2 + 2 G^2
     \langle N_g^2 \rangle}} \approx {{2 ~G^6 \langle N_g^2
     \rangle} \over {S_g^4 + 2 G^2 \langle N_g^2 \rangle}}
     ~~. \eqno(2.9)$$
This defines the residual uncertainty in the PSD due to noise which we
use to test the significance of deviations between two estimated PSDs.

If two bursts are lensed copies, we expect $\til {G}^2 - \til {H}^2
\approx 0$.  We measure the significance of  deviations from zero by
defining the $\chi^2$ statistic:
$$\chi_\nu^2 / \nu
     \equiv {{1}\over{N_f - 2}} \sum_f {{ (\til {G}^2 - \til {H}^2)^2 }
     \over {\langle ( \til {G}^2 - G^2 + H^2 - \til {H}^2 )^2 \rangle}}
     ~~,
     \eqno(2.10)$$
where the sum is over the $N_f$ frequency bins not
dominated by the noise (here defined as $\til {G}^2/\langle N_g^2
\rangle \ge 3$ {\it and} $\til {H}^2/\langle N_h^2 \rangle \ge 3$).  We
also define the degrees of freedom to be $\nu \equiv N_f - 2$, since we
subtract two degrees of freedom due to the two constraints provided by
the filters, $\Phi_g$ and $\Phi_h$.  We require at least three
frequency channels that meet our signal dominated criterion, and this
places a limit to how faint a burst to which we can apply this
statistic.

\smallskip
\centerline{\it b) Comparison via Cross Power Spectral Densities}
\smallskip

Just as we defined an optimal filter for the PSDs, we now define an
optimal filter, $\Phi_{gh}$, for the CPD, given by
$$\til {C} \equiv \til {GH^*} \equiv S_g S_h^* ~\Phi_{gh} ~~,
\eqno(2.11)$$
such that $\langle (\til {GH^*} - GH^* )^2 \rangle$ is
minimized. Following the same procedures as above, we can show that
$$\Phi_{gh} = {{G^2 H^2}\over{G^2 H^2 + G^2 \langle N_h^2 \rangle + H^2
     \langle N_g^2 \rangle + \langle N_g^2 \rangle \langle
     N_h^2 \rangle}} \approx {{G^2 H^2}\over{S_g^2 S_h^2}}
     ~~. \eqno(2.12)$$
Similar to the above, we take $G^2 H^2 \approx S_g^2 S_h^2 - \til
{G}^2  \langle N_h^2 \rangle - \til {H}^2 \langle N_g^2 \rangle -
\langle N_g^2 \rangle \langle N_h^2 \rangle$.

The CPD is a complex quantity, having both an amplitude and a
(typically) non-zero phase.    We write the
phase of the intrinsic signal as $2 \pi \tau_0(f) f$, and the phase
of the optimized signal as $2 \pi \tau_m(f) f$.
We write the optimized and intrinsic CPDs as
$$\eqalign{ \til {C} = \til {GH^*} &= C_m e^{i 2 \pi \tau_m(f) f} \cr
            C = GH^*  &= C_0 e^{i 2 \pi \tau_0 f}  ~~. }
\eqno(2.13)$$
If the intrinsic
signals are  scaled, time-shifted versions of each other, then
$\tau_0(f)$ is constant; however, $\tau_m(f)$ will vary due to noise.
We wish to determine how large a variation is expected in the
presence of noise under the assumption that $\tau_0$ is indeed a
constant.

If the light curves are lensed copies of each other,
we expect that $\til {C} - C \approx 0$.  In the limit of
small noise, we can also take $C_m \sim C_0$, and therefore the
squared difference between the measured and true transforms can be
written as
$${{(\til {C} - C)^2}\over{C^2}} \approx 4 \sin^2 {\pi {[\tau_m(f) -
     \tau_0] f}} \approx 4 \pi^2 [\tau_m(f) -\tau_0]^2 f^2
     ~~, \eqno(2.14)$$
which measures the phase difference between the optimized and
true CPDs. Unfortunately, we do
not know the true time delay, $\tau_0$, so instead  we calculate
the ``true" phase by taking $\tau_0$ to be the mean of the
measured delays, \ie $\tau_0 \approx \langle \tau_m(f) \rangle$.

We can quantify the dispersion of the measured CPD phase due to noise
from our knowledge of the statistical properties of our optimized
transforms.  Specifically, using equation (2.12) we can show
$${{\langle (\til {C} - C)^2 \rangle}\over{C^2}} = (1 - \Phi_{gh})
     ~~. \eqno(2.15)$$
With this as the measure of the uncertainty of the phase resulting
from noise, we define the $\chi^2$ statistic:
$$ \eqalign{\chi^2_\nu / \nu
     &\equiv {1 \over {N_f-2}} \sum_f {{(\til {C} - C)^2}\over
     {\langle (\til {C} - C)^2 \rangle}} ~
     \cr
     &= {4 \pi^2 \over {N_f-2}} \sum_f {{[\tau_m(f) -
     \langle \tau_m \rangle ]^2 f^2}
     \over {(1 - \Phi_{gh})}} ~~, } \eqno(2.16)$$
where as before we sum over the signal dominated frequency channels,
and $\nu = N_f - 2$ is the degrees of freedom (we subtract two
degrees of freedom for the constraints provided by $\Phi_{gh}$ and
$\langle \tau_m  \rangle$). Equations (2.10) and (2.16) together are
the statistical tests that we shall apply to real burst light
curves.

\bigskip
\centerline{3. Application to Simulated Burst Data}
\medskip

\centerline{a) Application to Scaled Bursts}
\smallskip

     We apply the statistics developed above to
the problem of distinguishing two similar but distinct burst profiles.
Following Wambsganss (1993), we choose burst GRB 910503 for our test.
This burst consists of two subbursts: a bright, spikey burst (which
we call subburst A) and a weaker, smoother burst (which we call
subburst B). Each subburst has a duration of roughly 10 s.  The
background subtracted burst light curves are shown in Figure 1.  We
have chosen this burst for the same reasons as Wambsganss; namely, this
burst is one of the brightest BATSE bursts and therefore should be
easily distinguishable. Most bursts, however, are much fainter;
therefore, one should check the possibility that subbursts A and B
could have been falsely identified as a lensing event if they were
fainter.  In addition, by choosing this event we can make direct
comparisons between our method and that of Wambsganss.

We follow the same methods that Wambsganss (1993) used to rescale the
burst light curves to lower amplitudes. We assume that the unscaled,
background subtracted burst profiles are the ``true" ({\it i.e.},
noiseless) burst profiles.  First, these profiles are shifted by
a random fraction of a data bin (each bin is 64 ms, and the photons are
assumed to be evenly distributed within each bin).  The burst is then
rescaled to a fainter level and the background is readded. Gaussian
noise is then added to the profiles, simulating the noisy signal of a
fainter burst, and the background is resubtracted.

We compare the bursts using information
provided by their PSDs and CPDs. In Figure 2, we plot the PSDs
(based upon a 256 point FFT, spanning a duration of 16.4 s) of
subbursts A and B for various rescalings.  As a consequence of our
PSD normalization, the amplitude of the PSD in the
signal dominated frequencies is independent of scaling (\cf Appendix
A).  The amplitude of the noise dominated channels, however, increases
by roughly a factor of 10 for each factor of 3 reduction of the
signal amplitude in the time domain. There are two other things to
notice here.  First, the PSD's of the two subbursts are intrinsically
very similar, approximately following a $1/f^2$ power law
(the form for a fast rise and exponential decay [FRED]). The
similarities are a consequence of the two bursts having comparable rise
and decay times.  Second, even for rescalings to much fainter levels,
the low frequency power is well defined and roughly unchanged.  This
gives hope that the method will be useful even for faint bursts.
In Figure 3 we plot the time delay [\ie $\tau_m(f)$] calculated from
the CPD of subburst A and B. We immediately notice that this time
delay is inconsistent with a constant, even at the lowest
frequencies.  As seen in Figure 3, the fourth lowest frequency bin,
corresponding to  $\sim 0.24$ Hz, has a time delay that is much
larger and of the opposite sign from the three lowest
frequency bins.

\vfill
\eject

\null
\baselineskip=12pt
\vfill
\noindent{{\bf Figure 1}:}  Background subtracted time profiles for
subbursts A (top) and B (bottom) of GRB 910503.  The solid line
correponds to photon count per 64 ms time bin, and the dashed line
to the burst signal to noise ratio for 2.048 s time bins.

\eject

\null
\vskip 8.0 true cm
\noindent{{\bf Figure 2}:}  Power spectral densities for the 256 point
FFTs of subbursts A (left) and B (right), for several rescalings of the
burst amplitude.  The PSD normalization is as defined in Appendix A.
Dotted lines are the expected amplitudes for the noise PSD when the
subbursts are scaled to 3\%, 10\%, 30\%, and 100\%.

\vfill
\noindent{{\bf Figure 3}:} Time delay (defined as the phase of the
cross power spectral density divided by Fourier frequency) between
subbursts A and B.  The solid line is the computed value, and dashed
lines are the $2~\sigma$ noise limits.

\eject

\baselineskip=14pt

We compare subbursts A and B scaled respectively to levels (32\%,
32\%), (8\%, 16\%), (4\%, 8\%), and (2\%, 8\%).  We also compare A to
itself using the rescalings 32\%, 8\%, 4\%, 2\%, and  B to itself using
the rescalings 32\%, 16\%, 8\%.  For each comparison we generate 1000
noisy light curve pairs, and we calculate for each pair the  degrees of
freedom ({\it i.e.}, number of frequency channels above noise minus 2) and the
corresponding reduced $\chi^2$ for the PSD and CPD statistics.
Histograms of the results are presented in Figures 4 and 5.

In Figure 4 for the PSD statistic, the top row
shows the reduced $\chi^2$ histograms for the A vs. B comparisons,
the middle row shows the histograms for the A vs. A comparisons, and
the bottom row shows the histograms for the B vs. B comparisons.
{}From left to right, the bursts are rescaled to fainter levels.
Each panel gives the number of
statistically well-defined simulations ({\it i.e.}, those with 1 or more
degrees of freedom), the mean degrees of freedom, $\bar \nu$, and the
mean reduced $\chi^2$, $\overline{\chi_\nu^2/\nu}$.  The means are
calculated only from the well-defined simulations.  Plotted
on top of the histograms are true reduced $\chi^2$ probability
distributions (\cf Abramowitz and Stegun 1972), with the degrees of
freedom taken to be $\bar \nu$ for that particular histogram.   These
distributions are normalized such that their integrated areas are
identical to those of the histograms.

It is obvious from visual inspection alone that when subbursts A and
B are not reduced below $\sim 10\%$ of their original flux, their
$\chi^2$'s are well above that expected from noise alone and they are
easily distinguishable by the PSD statistic.  In particular, most of
the area in the A vs. B histograms at (32\%, 32\%) and (8\%, 16\%)
fall far outside the true reduced $\chi^2$ distributions.  For the
comparison (4\%, 8\%), which Wambsganss (1993) considered too similar
to be unambiguously distinguished, the overlap between the reduced
$\chi^2$ distribution and the calculated histogram is indeed
substantial. Nevertheless, a fair fraction of the histogram lies
outside the distribution.  For the case (2\%, 8\%), which Wambsganss
termed ``impossible", it is indeed very difficult to distinguish the
reduced $\chi^2$ distribution from the PSD statistic histogram.

\baselineskip=12pt
\footnote{ }{\noindent {\bf Figure 4} {\it (following page)}:
Reduced $\chi^2$ histograms for the PSD statistic (equation [2.9])
based upon 1000 Monte Carlo simulations for each rescaling.  {\it Top
row:} From left to right, A vs. B at (32\%, 32\%), (8\%, 16\%), (4\%,
8\%), (2\%,8\%).  {\it Middle row:}  From left to right, A vs. itself
at 32\%, 8\%, 4\%, 2\%.  {\it Bottom row:} B vs. itself at 32\%, 16\%,
8\%. Each panel lists the number of simulations with well defined
statistics (\ie three or more frequency bins above the noise, \cf \S
2), the mean degrees of freedom ($\bar \nu$, the average number of
frequency bins above noise minus two), and the mean reduced $\chi^2$
($\overline {\chi^2/\nu}$).  Solid lines in each panel correspond a
true reduced $\chi^2$ distribution with $\bar \nu$ degrees of freedom
(\cf Abramowitz and Stegun 1972), normalized to have the same area as
the histogram.}

\baselineskip=14pt
The comparisons of A with itself and B with itself should show
agreement to within the noise.  These comparisons are useful in
determining to what extent the PSD statistic follows a true reduced
$\chi^2$ distribution.  From visual inspection, it apppears that
measured distributions with more degrees of freedom have
histograms that are closer to a true reduced $\chi^2$ distribution.  As
the degrees of freedom decrease (with the downscaling of the
subbursts), the histograms tend to be skewed to values greater than a
true reduced $\chi^2$ distribution.  In fact, the histograms for A vs.
A at  2\% and B vs. B at 8\% are quite similar to the A vs. B at (2\%,
8\%) histogram. This suggests that the latter comparison is
mostly consistent with identical light curves differing only by the
noise. For distributions with few degrees of freedom, we are hesitant
to assign a firm ``significance value" to our measured $\chi^2$ since
the probability distribution does not follow a true $\chi^2$
distribution.   Any such numbers should be indicative of a trend,  but
their significances are not directly quantifiable. For example, a $5
\sigma$ $\chi^2$ might be in reality $4 - 6 \sigma$, but large
$\chi^2$'s are definitely significant.

\pageno=13

Table 1 quantifies the above discussion.  In it we present
the percentage of (statistically well-defined) simulations that are
statistically different at
$\ge 1~\sigma$, $\ge 2~\sigma$, and $\ge 3~\sigma$ confidence levels,
as calculated from a true reduced $\chi^2$ distribution.  These
confidence levels correspond to the most extreme $32\%$, $4.5\%$, or
$0.3\%$ of the distribution, respectively.  For A vs. B at (32\%,
32\%), 100\% of the simulations are different with $3~\sigma$
confidence.  For A vs. B at (8\%, 16\%), 81\% are different with
$3~\sigma$ confidence.  The percentage drops markedly for A vs. B at
(4\%, 8\%); however, 26\% of the defined simulations
($19\%$ of all the simulations) are different with $3~\sigma$
confidence.  This is greater, by a factor of 2.5, than the most skewed
A vs. A or B vs. B comparisons, indicating greater differences than
can be understood from noise alone. This suggests that the statistic,
though becoming difficult to calibrate, is still useful at such a low
rescaling.  The A vs. B at (2\%, 8\%)
simulations are defined only 19\% of the time, and only $8\%$ of these
simulations are greater than $3~\sigma$ significant.  This percentage
is  comparable to the A vs. A at 2\%  and B vs. B at 8\% comparisons,
and therefore most of the greater than $3~\sigma$ results cannot be
considered significant.

With an average of 16.0 and 7.9 degrees of freedom respectively, the
A vs. A at 32\% and A vs. A at 8\% simulations are the
closest to a true reduced $\chi^2$ distribution.  All the lower
amplitude simulations show pronounced excesses of large deviations,
with 15-35 times too many $3~\sigma$ events.  In light of this,
it would be prudent to regard a $3~\sigma$ deviation (as measured by
a true $\chi^2$ distribution) in the PSD statistic as the lower bound
for a ``significant deviation" for distinguishing two bursts.

In Figure 5 we present histograms for
the CPD statistic.  As before, the comparisons between A and B show
that they are fairly distinct when the subbursts are brighter than
$\sim 10\%$ their original flux.  At lower levels the comparisons
become more consistent with deviations due to noise only.
Unlike the PSD statistic, however, the self comparison (A vs. A and B
vs. B) histograms are significantly narrower than a true
reduced $\chi^2$ distribution.  As with the previous statistic, fewer
degrees of freedom broadens the histogram, so that in this case the
CPD histogram becomes more similar to a true reduced $\chi^2$
distribution.  These graphical results are quantified in Table 2,
which is the analogue of Table 1 for the CPD statistic.

As before, we define the 1$~\sigma$, 2$~\sigma$, and 3$~\sigma$
confidence levels as the most extreme 38\%, 4.5\%, and 0.3\% points of
the true reduced $\chi^2$ distribution.  {\it All} the A vs. B
comparisons show more 1$~\sigma$, 2$~\sigma$, and $3~\sigma$ deviations
than either the self comparisons or the true reduced $\chi^2$
distributions.  This is true of even the ``impossible" A vs. B at
(2\%, 8\%) (for which 5.5\% of the {\it total} simulations were
distinguishable at $\ge 2~\sigma$, and  2\% of the {\it total}
simulations distinguishable at $\ge 3~\sigma$--- making it merely
exceedingly difficult to distinguish).  If we adopt the criteria of
requiring a $\ge 2~\sigma$ significance in the CPD statistic, then
nearly 20\% of the A vs. B at (4\%, 8\%) simulations are
``distinguishable".  This is comparable to what was found for the PSD
statistic with a $3~\sigma$ criterion.

At this point we should note that the PSD and CPD statistics are
independent, since noise in the Fourier phase is uncorrelated with
noise in the Fourier amplitude.  This is demonstrated in Figure 6
where we plot the two statistics against each other for A vs. A at
4\%  and B vs. B at 8\%.  The absence of correlation justifies using
the PSD and CPD statistics jointly as even more
powerful discriminants for distinguishing bursts. If we adopt the
criterion that a burst is distinguished if {\it either} the PSD
statistic is $\ge 3~\sigma$ significant or the CPD statistic is $\ge
2~\sigma$ significant, then for the A vs. B at (4\%, 8\%), 47\% of the
well-defined simulations are distinguished (34\% of all the
simulations). This is to be compared to 11\% of the well-defined A vs.
A at 4\% and 12\% of the B vs. B at 8\%  simulations being falsely
distinguished.

It is not surprising that our method is slightly more successful at
distinguishing bursts than Wambsganss's method.  In his method, {\it
all} time bins in a light curve are considered, whether they are
dominated by noise or signal. As the bursts are scaled downward, more
and more time bins become noise dominated until the statistic itself
becomes noise dominated.  As Wambsganss noted in his paper, if he
rebinned the data by combining 4 time bins at a time, the sensitivity
of his statistic improved.  This is not surprising as the signal to
noise ratio increases in each new bin, and in some ways this is
analogous to using the lower frequency bins in the Fourier domain.  As
an extreme case, each burst can be rebinned into a single time bin, and
then a best fit scale factor and time delay can be computed.  Each
burst then can be rebinned into two time bins, and the constancy---
with respect to the noise levels--- of the scale factor and time delay
can be checked. The procedure can be
repeated for four, eight, sixteen, etc., time bins, until the
bins become noise dominated. However,
Wambsganss only used the finest resolution
time bins and thus diluted his error measure with more noise than
necessary.  Our procedure is essentially equivalent to a rebinning that
uses the finest time bins {\it allowed by our signal to noise
criteria}, hence enhancing the discriminatory power of our statistics.

\baselineskip=12pt
\footnote{ }{\noindent {\bf Figure 5} {\it (following page)}:
Same as Figure 4, except for the CPD statistic.}

\vfill
\eject

\null
\pageno=17
\baselineskip=12pt
\vfill
\noindent{{\bf Figure 6}:}  The reduced $\chi^2$ for the CPD statistic
[$(\chi^2_\nu/\nu)_{CPD}$] vs. the reduced $\chi^2$ for the PSD
statistic [$(\chi^2_\nu/\nu)_{PSD}$].  {\it Top:} Subburst A vs.
itself at 4\%.  {\it Bottom:} Subburst B vs. itself at 8\%.  Note the
lack of correlation.
\eject

\baselineskip=14pt
{\centerline{\it b) Interpretation in Terms of Signal to Noise
Ratios}}
\smallskip

To understand what the above results mean for real burst data, it is
necessary to discuss their interpretation in
terms of the BATSE signal to noise ratio (\cf
Fishman et al. 1992).  BATSE consists of eight detectors, essentially
forming the eight faces of an octahedron, hence a burst will
illuminate two to four detectors.  Each detector is capable of
observing a burst in a variety of spectral ranges; however,the
coarsest spectral resolution (which is used for triggering the
detectors) consists of only four energy channels: 20-50 keV, 50-100
keV, 100-300 keV, and 300+ keV (we will refer to these as channels 1-4,
respectively). BATSE triggers when the signal in the {\it second} most
illuminated detector exceeds the background noise level by a factor of
5.5.  The signal and noise are measured from the sum of channels 2
and 3 only.  BATSE can trigger on timescales of 64 ms, 256 ms, or 1024
ms, the latter being  the most sensitive for sufficiently long and
smooth bursts.  Below, we shall consider only the 1024 ms trigger
threshold, and any burst whose peak flux (in 1024 ms, between 50-300
keV, in one detector) is at the detection threshold will be said to
have $C_{max}/C_{min} = 1$. $C_{max}$ and $C_{min}$ are, respectively,
the maximum detected count rate and the minimum detecable count rate.
Any burst with an intrinsic flux such that $C_{max}/C_{min} = 1$ has a
$\sim 50\%$ probability of triggering BATSE. (Noise fluctuations will
reduce the observed flux below the threshold roughly half the time.)
In order for the burst to trigger BATSE 100\% of the time, the
intrinsic flux must correspond to $C_{max}/C_{min} \approx 1.5$ (\cf
BATSE trigger tables, Fishman et al. 1992).

The particular light curve used by us and Wambsganss (1993) is the
summed 20-50 keV (channel 1) data of four BATSE detectors, making
direct comparison between the burst rescalings used above and the true
BATSE detection threshold difficult. In order to make these
comparisons, we shall pretend that the single energy channel light
curve that we used actually corresponds to a hypothetical complete
light curve pieced together from {\it all four} available energy
channels and two to four on-burst detectors. The signal to noise
ratio of the light curve in the energy channels that trigger BATSE
are typically less than the signal to noise ratio in the full
light curve (channels 1-4 summed).  We therefore must determine what
fraction of our channel 1 light curves would correspond to
the BATSE trigger if they were in reality
complete light curves (2-4 detectors, channels 1-4, summed).  To do
this, we use the full light curve ({\it i.e.}, summed over all energy
channels) of GRB 910503 as an example of a ``typical" gamma ray
burst.  We use this burst to determine the relationship between the
signal to noise ratio in the trigger channels 2-3, and the (larger)
signal to noise ratio in the full burst.  We assume that our channel 1
light curves would show the same relationship if they were in reality
complete light curves.  We use these estimates as a means of assigning
to the rescalings we performed above a $C_{max}/C_{min}$, which here
shall {\it always} refer to the peak count rate (over 1024 ms) divided
by the threshold in the 50-300 keV band of the second most illuminated
detector.

Assuming 144 photons per 64 ms as the ``canonical" background rate
in the 50-300 keV energy band, the minimum detectable $5.5 ~\sigma$
signal will be 264 photons per 1024 ms {\it in one detector} ($264 =
5.5 \sqrt{1024 ms/64 ms \times 144}$).  Realistically, we will have at
least one more detector, and an additional 40\% photon flux due to the
20-50 keV and 300+ keV band passes.  (The 40\% increase is based upon
the spectrum of GRB 910503.)  The minimum summed signal is therefore
740 photons per 1024 ms.  As our background, we take a typical value
to be 300 photons per 64 ms per detector (again, based upon the
spectrum of the background of GRB 910503), which yields a Poisson noise
level of 98 photons per 1024 ms (2 detectors, channels 1-4). The {\it
minimum} signal to noise ratio in the summed burst is therefore
$740/98 = 7.6$.  For the chanel 1 light curves that we
used, the peak signal to noise  ratio on the 1024 ms time scale is
approximately $32300 X_A/ 84 \approx 385 ~X_A$ for subburst A and
$13300 X_B / 84 \approx 158 ~X_B$ for subburst B, where $X_A$ and $X_B$
are the factors by which the subbursts are scaled (84 photons per 1024
ms is the noise level in channel 1, and 32300 photons per 1024 ms and
13300 photons per 1024 ms are the peak count rates in this channel for
subbursts A and B respectively).  In terms of $C_{max}/C_{min}$,
subbursts A and B would have $C_{max}/C_{min} \sim 50.7~X_A$ and
$C_{max}/C_{min} \sim 20.8~X_B$, respectively, if these channel 1
light curves were in reality complete light curves (2 detectors,
channels 1-4).

Comparing these results to our rescalings, we see that
$C_{max}/C_{min} = 1$ (50\% probability of detection) corresponds to A
and B at (2\%, 4.8\%), and $C_{max}/C_{min} = 1.5$ ($\sim$100\%
probability of detection) corresponds to A and B at (3\%, 7.2\%).
Note that this is for only two detectors.  If the burst is seen in
three detectors with roughly equal flux, then the signal to noise
ratio in the summed burst light curve increases by a factor of
$\sqrt{3/2}$.  For this case, the 100\% probability of detection
($C_{max}/C_{min} \approx 1.5$) corresponds to A and B at (3.6\%,
8.8\%).  For four equally illuminated detectors, the $\sim$100\%
probability of detection corresponds to A and B at (6\%, 9.6\%).
Wambsganss's interpretation of A  vs. B scaled at (2\%, 8\%)
being at the BATSE detection threshold is something of a worst case
scenario.  A likley lower limit to bursts that would trigger BATSE
corresponds to A vs. B at (4\%, 8\%), which as we
demonstrated before yields 34\% of the comparisons being
distinguishable by {\it at least one} of our statisitical tests.
Depending upon the energy spectra of individual burst light curves and
the background, it may be possible to choose more
optimal combinations of detectors and energy channels so as to achieve
even greater signal to noise ratios.

We saw above that for A vs. B at (8\%, 16\%), each individual
statistic was capable of distinguishing $\sim 80\%$ of the pairs.
Therefore, $\sim 96\%$ of the pairs are distinguishable by at least
one of the tests.  This is the level of rescaling at which we have
strong confidence in our tests.  In terms of $C_{max}/C_{min}$, this
rescaling corresponds to: $C_{max}/C_{min} \approx 4.1$ for A,
$C_{max}/C_{min} \approx 3.3$ for B (two detectors); $C_{max}/C_{min}
\approx 3.3$ for A, $C_{max}/C_{min} \approx 2.7$ for B (three
detectors); $C_{max}/C_{min} \approx 2.0$ for A, $C_{max}/C_{min}
\approx 1.7$ for B (four detectors). We therefore expect that when
$C_{max}/C_{min} \aproxlt 3$ for two intrinsically similar bursts, they
will not be consistently distinguishable.  For the 211 bursts from
the burst catalogue (Fishman et al. 1992) that have a known
$C_{max}/C_{min}$ on the 1024 ms timescale, 59\% have $C_{max}/C_{min}
< 3$ and 45\% have $C_{max}/C_{min} < 2$.  Thus, roughly half of all
bursts will not be consistently distinguishable, within the noise, from
another similar, but different, light curve.

We should note that these limits apply to {\it intrinsically} similar
bursts.  The light curves of subbursts A and B are very similar on
the coarsest time scales.  Returning to Figures 2 and 3, we see that
for the three of the four lowest frequencies (0.06 Hz, 0.12 Hz, 0.24
Hz) the PSD amplitudes for the two subbursts are nearly identical.
For the three lowest frequencies, the CPD time delay is small and near
zero.  This is just a statement that the coarsest features ({\it
i.e.}, rise and decay times, broad spikes, etc.) are very similar from
subburst A to B.  These subbursts will appear to be statistically
identical for burst rescalings wherein only the few lowest frequency
bins are above the noise. The only way to determine how common
such similar but distinct pairs are would be to compare pairs of
bursts whose angular separations are much
larger than the positional errors.  If indistinguishable pairs
are uncommon, then we would have greater confidence that an
indistinguishable pair within the same error box was due to lensing.

Our statistics also require at least three frequency bins above the
noise. The shortest Fourier transform that yields three or more
distinct frequency channels is the 8 point FFT (we only consider
transforms whose length are a power of 2, \cf Press et al. 1992).
This FFT yields a PSD at the frequencies: $(1/4) ~f_{ny}$, $(1/2)
{}~f_{ny}$, $(3/4) ~f_{ny}$, $f_{ny}$; where $f_{ny}$ is the Nyquist frequency.
 The
Nyquist frequncy is the the highest frequency calculable by an FFT,
and for an 8 point FFT it is given (in Hz) by $f_{ny} = 4/T$, where $T$ is
the duration of the data.  We expect that in order to have a
well-defined FFT, half of our time bins must be above the noise.  In
Figure 1, we show the signal to noise ratio for subbursts A and B
rebinned into 8 time bins. At 100\% scaling, the bursts are well above
the noise.  However, when A is scaled to 4\%, three time bins are
clearly below the noise, two are marginal at 4-6 times the noise, and
only three are well above the noise.  When A is scaled to 2\%, four
time bins have signal to noise less than three (our usual criteria for
a detectable signal), one is marginal at 3 times the noise, and only
three are well above the noise.  This reduction of signal to noise
is reflected in the fact that only  73\% of the A vs. B at (4\%, 8\%)
comparisons have well-defined statistics, and only 19\% of the A
vs. B at (2\%, 8\%) have well-defined statistics.  Again, the only
way to know how often the statistic will be defined in practice is
to compare a large number of real burst pairs.  In the next section,
we do this for a small number of lensing event candidates.

\bigskip
\centerline{4. Application to Real Burst Pairs}
\medskip

As discussed above, in order to fully calibrate our
statistic, or any other statistic that attempts to distinguish burst
profiles, one should compare all possible burst pairs and see how
many of these pairs are indistinguishable.  Pairs that are well
outside of each others positional error boxes cannot be lensing
events. Therefore the fraction of such events that are
indistinguishable would give a good indication of the number
of ``false positives" among lensing candidates.  With 260 bursts in the
BATSE catalogue, and hence 33670 possible pairs, this is a daunting
task.  One could restrict the comparison to bursts of similar
duration, such as only using pairs of bursts whose durations are
within a factor of two of each other.  A convenient measure of a
burst's duration is given by its $t_{50}$ and $t_{90}$ times, which
are defined as the time intervals during which the burst's integrated
photon counts go from 25\% to 75\% and 5\% to 95\%, respectively, of
the total photon counts (Fishman et al. 1992).  With this
restriction, which we adopt below, there are 1648 possible pairs,
which is still a rather large number of comparisons.

Instead, we apply our statistic to the subset of the 1648 pairs
where the positional error boxes of the pair overlap.  Each
burst's positional error is taken to be $4^\circ$ (the BATSE
systematic error) added in quadruture with its positional error as
listed in the BATSE catalog (Fishman et al., 1992).  The positional
errors for the two bursts are then added in quadruture, giving us a
total angular error between the two.  Only 25 pairs of bursts with
similar durations are separated by an angle less than this error.
Table 3 lists these 25 pairs (45 separate bursts).  They are potential
lensing candidates.  In Table 3, bursts are identified by their BATSE
trigger number (Fishman et al. 1992).  In addition we include two
pairs--- bursts (235, 1447) and (493, 1430)--- that did not show up in
our search of the BATSE catalogs.  Burst 1430 is not listed in the
duration tables, and the bursts (235, 1447) are farther apart than
their positional errors.  Both pairs, however, are listed in Nemiroff
et al. (1993a) as potential lensing candidates.  We include these pairs
in order to compare our statistics to those of Nemiroff et al..

We have applied both of our statistics to each of the 27 pairs listed
in Table 3.  All the burst light curves are
constructed from the sum of {\it all} energy channels and {\it all}
on-burst detectors.  We chose a length for our Fourier transforms
such that the number of 64 ms time bins was the smallest
power of 2 (\cf Press et al. 1993) such that the duration of the
transform was longer than the $t_{90}$ time of the longest burst.
The background for each burst was assumed to be a constant and was
calculated by averaging 50 points beyond the end of each transform.
This background was subtracted before performing the Fourier
transforms.  Several of the light curves had levels at the beginning
of the data set that were inconsistent with the background at the end
of the data set.  This could be due to a time-varying background, or
the beginning of the burst may be absent from the data set.  (BATSE
stores only 2.048 s before each trigger, so bursts that have
durations longer than this before triggering may have portions of the
light curve missing.)  These bursts are noted with a superscript $b$
in Table 3.

Our statistics are applicable to 19 out of the 27 pairs, and {\it
all} of these pairs are distinguishable. The most similar pair
are bursts (235, 1447), which, although indistinguishable by the CPD
statistic, have a formal probability of only $10^{-3}$ of being
identical to within noise by the PSD statistic. This pair was
indistinguishable by Nemiroff et al. (1993a), but was not seriously
considered to be a lensing event due to the large angular separation
between the two and the fact that one of the bursts (235) was found to
have a position-consistent precursor in other BATSE data.  Pair (493,
1430), which Nemiroff et al. (1993a) calculated as having only a
probability of $10^{-2}$ of being statistically identical, is
distinguished with even greater confidence by our statistics.
Although this pair is indistinguishable by the CPD statistic,
the PSD statistic gives a probability of only $6 \times 10^{-6}$ of
the pair being statistically identical.  All of the remaining 17 pairs
to which we were able to apply our statistics are distinguishable {\it
with greater than} $3~\sigma$ confidence.

We were not able to apply our statistics to 8 pairs.  One of these
pairs, bursts (942, 1298), was too noisy to apply our statistics,
having only one frequency channel (yielding -1 degrees of freedom)
above the noise. The remaining 7 pairs were too short to
compare.  At least one burst within each of these pairs had three or
fewer time bins more than $3~\sigma$ above the noise.  As discussed
above, we require a light curve with at least 4 time bins clearly
above the noise, to which we can apply an 8 (or more) point FFT.  It
is not likely that {\it any} statistic based upon the temporal profile
alone will be able to adequately distinguish bursts with only 3 time
bins above the noise.   Fortunately, 90\% of the bursts have $t_{90}$
times greater than 0.256 ms (i.e. four 64 ms time bins).

As an example of the difficulty of distinguishing short bursts, we
show pair (138, 444), which is {\it completely} consistent with
lensing.  Not only are the light curves nearly identical, the bursts
are also achromatic, as is shown in Figures 7a,b.  In Figure 7a
we plot the two bursts, summed over all detectors and energy channels,
on top of one another.  Burst 444 is scaled so that its peak is
identical to the peak of burst 138.  In Figure 7b we show the
color-color diagrams for each time bin that is at least $2~\sigma$
above the noise.  (We loosen our signal to noise criteria somewhat so
as to have more time bins to compare.)  The colors are defined as the
ratios of energy (channel 2)/(channel 1) and (channel 3)/(channel
2).  The overlapping time bins have colors that are
consistent within the noise.  The brighter of
the two bursts has a third time bin with well defined colors, but its
counterpart does not. Being that the brighter burst arrives {\it after}
the fainter burst, this pair cannot be produced by a circular lens
(\cf Narayan \& Wallington 1992), but could be consistent with an
elliptical lens. We emphasize, however, that this pair is only
consistent with lensing because of a {\it lack} of information, not
because of abundant similarities.  Thus, we do not consider it a
lensing candidate.  This example points out the difficulties of being
certain that there are no lensing events in the burst catalogue,
except for the subset with durations $\aproxgt 0.3$ s.

\bigskip
\centerline{5. Summary}
\medskip

If lensed gamma ray burst light curves are to be used to demonstrate
the cosmological origin of GRBs, we must be able to identify such
events. As Wambsganss (1993) has pointed out, such events may be
difficult to distinguish from two noisy light curves that
\vfill
\eject
\null
\baselineskip=12pt
\vfill
\medskip
\noindent{{\bf Figure 7a,b}:} {\it a) Top:} Background subtracted
light curves for BATSE triggers 138 and 444.  Burst 444 is scaled to
have the same peak amplitude as burst 138.  Numbered points correspond
to points in the color-color diagram to the right.  {\it b) Bottom:}
Color-color diagram for bursts 138 and 444.  The vertical axis color
corresponds to BATSE energy (channel 1)/(channel 2) [ (20-50
keV)/(50-100 keV) ].  The horizontal axis color corresponds to BATSE
energy (channel 2)/(channel 3) [ (50-100 keV)/(100-300 keV) ].  Error
bars represent $1~\sigma$ uncertainties.
\eject
\noindent{are}
similar but distinct.  In this work, we developed two statistical
tests that are capable of distinguishing noisy light curves.  These
tests used Fourier transforms to compare bursts.  Comparisons were
made at frequency channels that are not noise dominated.  If the light
curves had, to within the noise, comparable Fourier amplitudes and
phases consistent with a constant time delay, they were said to be
statistically identical.

We then applied these statistics to two subbursts of GRB 910503, the
burst that Wambsganss (1993) considered.  We showed that as long as
the subbursts were not scaled to $\aproxlt 10\%$ of their original
amplitude, they were distinguishable.  Just as Wambsganss (1993)
noted, these bursts became harder to distinguish as they were further
reduced.  However, our results were not as pessimistic, since we found
that bursts that were only 50\% above the BATSE detection threshold
were still distinguishable one third of the time. In addition to our
tests being slightly more successful than Wambsganss (1993), they are
more quantitative in that they give a significance
value of the burst differences and an indication of the information
content of the comparison (through the degrees of freedom, the
number of frequency channels above noise minus 2).  We also
interpreted our results in terms of the BATSE detection thresholds.
These intrinsically similar subbursts were nearly 100\%
distinguishable for $C_{max}/C_{min} \approx 3$ ({\it i.e.}, three
times the detection threshold), 34\% distinguishable for
$C_{max}/C_{min} \approx 1.5$, and 5\% distinguishable for
$C_{max}/C_{min} \approx 1$.

We then applied our statistics to 27 lensing candidate burst pairs,
19 of which were found to be easily distinguishable.  Of the
remaining 8 pairs, one was too noisy to apply our statistics to,
and the others were too short, having fewer than four time bins
above the noise.  By this duration criterion, our statistic should
be applicable to more than 90\% of all bursts.  These calculations
could be improved in several ways.  In addition to the 64 ms time
binning, BATSE ``time tags" photon arrival times with a precision of
2 ms (Fishman et al. 1992).  Thus, the brighter short duration
bursts may be able to be examined at a higher time resolution,
allowing more detailed comparisons. Our statistics being
inapplicable to 10\% of all bursts therefore represents a worst case.
We have also taken the simplest approximation to the burst background,
that is a constant.  More detailed modelling may allow us to
distinguish the bursts labelled  $b$ in Table 3 with even greater
confidence.  Finally, we have not checked [with the exception of
bursts (138, 444)] that the burst pairs are achromatic.  This
undoubtedly will only enhance our ability to distinguish light
curves (perhaps even dramatically, according to some preliminary
tests [Nemiroff 1994b]).

We are less pessimistic than Wambsganss (1993), and believe that a
substantial fraction of the observed bursts should be
distinguishable from one another.  The best test of the sensitivity
of our statistics (or any statistic that attempts to distinguish GRB
light curves) would be to apply it to all burst pairs that {\it do
not} have overlapping error boxes.  Due to the enormity of this task
(over one thousand pairs), we have not attempted such a calibration.
The question then remains: how many unidentified lensed pairs should
exist in the BATSE data?  Recent results of Grossman \& Nowak
(1994) indicate that if the only lenses are galaxies, then it is
very unlikely that there are or will be {\it any} events observed
by BATSE.  However, other possibilities, such as lensing by massive
black holes or compact dark matter, may produce observable events
(Mao 1992, Blaes \& Webster 1992).  Even if a lensing event is
unlikely, the implications of one's discovery would be profound.  The
search is therefore worthwhile, and the techniques described above can
aid in this search.

\baselineskip=12pt
\medskip
\centerline{Acknowledgements}
\smallskip

The authors would like to thank Dr. Thomas Aldcroft, Dr. Rich Epstien,
Dr. Nick Ruggiero, Dr. Chris Schrader,  but most especially Dr. Ed
Fenimore for various help in obtaining and reading the burst data
files.  Many of the files we used we were supplied directly by Dr.
Fenimore, and for this we are eternally grateful. We also would like
to acknowledge useful discussions with Prof. Omer Blaes, Dr. David
Syer, and Dr. Brian Vaughan (who provided much insight into the
Fourier analysis techniques).  MAN would like to thank the Aspen
Center for Physics for its hospitality during the ``Gravitational
Problems in Relativistic Astrophysics" workshop, where part of this
work was started.  This research has made use of data obtained through
the Compton Observatory Science Support Center GOF account, provided
by the NASA-Goddard Space Flight Center.

\baselineskip=14pt
\bigskip
\centerline{Appendix A. Fourier Analysis in the Presence of Noise}
\medskip

The light curves that BATSE measures will be comprised of an
intrinsic signal (which we assume to be scaled by lensing),
and background, with noise superimposed on top of these components.  In
this Appendix, we compute the statistical properties of noise in the
Fourier domain, and we define a normaliztion for our Fourier
transforms such that the Fourier amplitudes are independent, to within
the noise, of any uniform scaling of the intrinsic signal.  Detailed
derivations of the probability distribution of power spectral
densities (PSDs, {\it i.e.}, squared Fourier amplitudes) in the
presence of Poisson noise can be found in Leahy {\it et al.} (1983)
and Groth (1975). These references use slightly different PSD
normalizations (wherein the level of the noise is fixed at a constant
value). Here we derive the statistics for our choice of normalization.

\smallskip
\centerline{\it The Noise PSD}
\smallskip

We have an intrinsic signal which we call $x^I(t)$ (for our
purposes, photon counts per second).  Assuming that we evenly
sample this signal at $N$ time bins, $t_k$, of
duration $\Dt$, we obtain a series of $N$ measured signals $\lxm{k}$
(photon counts).  The measured counts
will be different from the intrinsic counts, $\lxi{k} \approx x^I(t_k)
\Dt$, due to counting noise.  Data obtained from the BATSE experiment
is comprised of the sum of $1-4$ energy channels and $2-4$ ``on burst"
detectors; therefore, by the central limit theorem, the
measured signal is gaussian distributed about $n_k$ with variance
$n_k$, according to
$$P_k ( m_k ) = {{1}\over{\sqrt{2 \pi n_k}}} e^{-(m_k - n_k)^2/2 n_k}
      ~~. \eqno({\rm A.}1)$$

To study the data in the Fourier domain, we take the discrete
transform of the measured data.
$$\bxm = \smk \lxm{k} ~e^{2 \pi i f_j t_k} ~\Dt ~~, \eqno({\rm A.}2)$$
where $f_j = j/(N \Dt)$, and $j = -N/2, -N/2 + 1, \ldots, N/2$.
The transform of the intrinsic signal is
$$\bxi = \smk \lxi{k} ~e^{2 \pi i f_j t_k} ~\Dt ~~. \eqno({\rm A.}3)$$
The PSD of the noise is simply the difference between the true and
measured PSD, $\bxm^2 - \bxi^2 ~~$.
Expanding the PSDs, we find for the measured signal
$$\bxm^2 = \smk \lxm{k}^2 \Dt^2 + \sum_{{k,l = 0}\atop{k
\ne l}}^{N-1} \lxm{k} ~\lxm{l} ~e^{2 \pi i f_j (t_k - t_l)} ~\Dt^2
{}~~, \eqno({\rm A.}4)$$
and likewise for the intrinsic signal.
Averaging the measured signal over the probability distribution,
keeping in mind that from time bin to time bin the noise
is uncorrelated, we find
$$\eqalign{\langle {\bxm^2} \rangle &=
     \left [\int \prod_{k} dm_k ~ P_k(\lxm{k}) \right ]
     \bxm^2 \cr
     &=  \smk \left [ \lxi{k}^2 + \lxi{k}
     \right ] ~\Dt^2 + \sum_{{k,l = 0}\atop{k
     \ne l}}^{N-1} \lxi{k} ~\lxi{l} ~e^{2 \pi i f_j (t_k - t_l)}
     ~\Dt^2 ~~.} \eqno({\rm A.}5)$$
The first term of the first summation and the second
summation together are $\bxi^2$, so that
$$\langle {\bxm^2} \rangle - {\bxi^2} = \smk \lxi{k} ~\Dt^2 =
      \langle \smk \lxm{k} ~\Dt^2 \rangle ~~.
     \eqno({\rm A.}6)$$
This difference is the mean power spectral density due to noise, and
is simply $N_\gamma \Dt^2$ where
$$N_\gamma  = \smk \lxi{k} = \langle \smk \lxm{k} \rangle ~~,
     \eqno({\rm A.}7)$$
is the total photon count.  Note that the mean PSD due to noise is
constant, independent of frequency.

\smallskip
\centerline{\it PSD Normalization}
\smallskip

We choose a one-sided PSD normalization  (\cf Press et al. 1992) such
that the intrinsic power spectral density is given by
$$\bxip \equiv 2 ~{{N~ \bxi^2}\over{{N_{\gamma}}^2 \Dt}} ~~,
     \eqno({\rm A.}8)$$
where we recall that $N$ is the number of time or frequency bins.
With this normalization, summing the PSD over frequency yields the
mean square signal divided by the square of the mean signal, $\langle
n_k^2 \rangle / \langle n_k \rangle^2$.  This PSD is therefore
independent of any uniform scaling of the {\it total} signal, source
plus background.

In the above, $\lxi{k}$ is the intrinsic
photon count, which consists of both a signal from a source, $n^s_k$,
and the  signal from the background, $b_k$.  We wish to modify our
normalization so that the PSD is independent of any uniform scaling
of the {\it source} signal alone.  This is done simply by replacing
$N_\gamma$ in $({\rm A}.8)$ with
$$ N_s = \smk n_k - b_k  ~=~ \langle \smk  \lxm{k} -
     b_k  \rangle ~~, \eqno({\rm A.}9)$$
where $N_s$ is the {\it source} photon count.  (We subtract the
background from the signal before performing the transform.)

The noise level in our PSD will still be based upon the sum of the
source and the background. Combining $({\rm A}.7)$ with our
normalization gives the noise PSD as
 $$ 2 \Dt {{\langle \lxi{k} \rangle}\over {{\langle n_k -
     b_k   \rangle}^2}} ~~. \eqno({\rm A.}10)$$
This is the expression that we use to determine our noise level when
calculating our optimal filters (\S 2) and for determining when the
signal to noise level is  $\ge 3$.  When the background count rate,
$b_k$, dominates the total count rate, $n_k = n^s_k + b_k$, then a
factor of ten decrease in the  source count rate leads to a factor of
one hundred increase in the level of the noise PSD. This behavior is
seen in Figure 2.

\vfill
\eject
\baselineskip=12pt

\bigskip
\centerline{REFERENCES}
\bigskip

\bookpl{Abramowitz, M., \& Stegun, I.~A.}{1972}{Handbook of
Mathematical Functions}{New York}{Dover}

\noindent\hangindent=20pt\hangafter=1{
Blaes, O.~M. ~1994, ApJS, in press}

\apj{Blaes, O.~M., \& Webster, R.~L.}{1992}{391}{L63}

\bookpl{Davenport, W.~B., Jr., \& Root, W.~L.}{1958}{An Introduction
to the Theory of Random Signals and Noise}{New
York}{McGraw-Hill}

\noindent\hangindent=20pt\hangafter=1{
Fenimore, E.~E., et al. ~1993, Nature, 366, 40}

\noindent\hangindent=20pt\hangafter=1{
Fishman, G.~J., et al. ~1992, BATSE Burst Catalog, GROSSC}

\apjsup{Groth, E.~J.}{1975}{29}{285}

\apj{Leahy, D.~A., {et al.}}{1983}{266}{160}

\apj{Gould, A.}{1992}{386}{L5}

\noindent\hangindent=20pt\hangafter=1{Grossman, S.~A., \& Nowak,
M.~A. ~1994, ApJ,  submitted}

\noindent\hangindent=20pt\hangafter=1{
Hartmann, D.~H. ~1994, to be published in \lq\lq High Energy
Astrophysics", ed. J. Matthews (World Scientific)}

\apj{Mao, S.}{1992}{389}{L41}

\apj{Mao, S.}{1993}{402}{382}

\apj{Narayan, R., \& Wallington, S.}{1992}{399}{368}

\book{Nemiroff, R.~J., et al.}{1993a}{Proceedings of the Compton
Gamma-Ray Observatory Conference}{New York}{American Institute of
Physics}{974}

\apj{Nemiroff, R.~J., et al.}{1993b}{414}{36}

\noindent\hangindent=20pt\hangafter=1{
Nemiroff, R.~J. ~1994a, Comm. on Astr., to be published }

\noindent\hangindent=20pt\hangafter=1{
Nemiroff, R.~J. ~1994b, private communication}

\apj{Paczy\'nski, B.}{1986}{308}{L43}

\apj{Paczy\'nski, B.}{1987}{317}{L51}

\noindent\hangindent=20pt\hangafter=1{
Press, W., et al. ~1992, Numerical Recipes (Cambridge: Cambridge
University Press)}

\apj{Wambsganss, J.}{1993}{406}{29}

\noindent\hangindent=20pt\hangafter=1{
Wang, J.~L.~ 1994, Comm. on Astr., 17, in press}

\bookpl{Whalen, A.~D.}{1971}{Detection of Signals in Noise}{New
York}{Academic Press}

\vfill
\eject

\baselineskip=14pt

\def\horline{\multispan{10}\hrulefill \cr}

\centerline{\bf Table 1: Results of Power Spectra Simulations}
\bigskip

\centerline{\vbox{\offinterlineskip{\halign{\strut
{}~~~ # \hfill ~~ &~  # \hfill ~~ & \quad # & \hfill # ~~ &~~~ \hfill #
    ~~  & \hfill #  &~ \hfill # ~~ & \hfill #  &~ \hfill # ~~
    & \hfill #  \cr
\horline
\noalign{\vskip3pt}
\horline
\noalign{\vskip2pt}
Burst a & Burst b && $\bar \nu$ & $\% ~>~ 1\sigma$ && $\% ~>~ 2\sigma$
&& $\% ~>~ 3\sigma$ & \cr
\noalign{\vskip2pt}
\horline
\noalign{\vskip2pt}
A at 32\% & B at 32\% && 4.5 & 100.0 && 100.0 && 100.0 & \cr
A at ~8\% & B at 16\% && 2.2 & 99.6 && 96.8 && 81.2 & \cr
A at ~4\% & B at ~8\% && 1.2 & 94.4 & (69.2) &
     61.7 & (45.2) & 25.6 & (18.8) \cr
A at ~2\% & B at ~8\% && 1.1 & 82.9 & (16.0) &
     37.3 & (7.2) & 7.8 & (1.5) \cr
\noalign{\vskip2pt}
\horline
\noalign{\vskip2pt}
A at 32\% & A at 32\% && 16.0 & 20.8 && 2.7 && 0.1 & \cr
A at ~8\% & A at ~8\% && 7.9 & 53.2 && 12.8 && 1.3 & \cr
A at ~4\% & A at ~4\% && 2.5 & 65.5 & (64.4) & 23.4 & (23.0)
     & 7.2 & (7.1) \cr
A at ~2\% & A at ~2\% && 1.1 & 75.2 & (8.8) & 28.2 & (3.3)
     & 9.4 & (1.1) \cr
\noalign{\vskip2pt}
\horline
\noalign{\vskip2pt}
B at 32\% & B at 32\% && 3.4 & 62.8 && 19.3 && 5.1 & \cr
B at 16\% & B at 16\% && 1.7 & 78.8 && 32.0 && 10.4 & \cr
B at ~8\% & B at ~8\% && 1.0 & 83.9 & (65.6) & 34.5 & (27.0)
     & 7.4 & (5.8) \cr
\noalign{\vskip2pt}
\horline
}}}}
\bigskip
\baselineskip=12pt
Table 1:  Percentage of simulations above a given confidence limit,
based on 1000 simulations of the PSD statistic. The 1$~\sigma$,
2$~\sigma$, and 3$~\sigma$ thresholds correspond to the 68\%, 95.5\%,
and 99.7\% levels for a reduced $\chi^2$ distribution with $\bar \nu$
degrees of freedom.  Percentages are based upon those simulations with
well defined statistics ($\le 1000$). Numbers in parentheses
are based on all 1000 simulations.

\baselineskip=14pt
\bigskip
\bigskip

\centerline{\bf Table 2: Results of Cross Power Spectra Simulations}
\bigskip

\centerline{\vbox{\offinterlineskip{\halign{\strut
{}~~~ # \hfill ~~ &~  # \hfill ~~ & \quad # & \hfill # ~~ &~~~ \hfill #
    ~~  & \hfill #  &~ \hfill # ~~ & \hfill #  &~ \hfill # ~~
    & \hfill #  \cr
\horline
\noalign{\vskip3pt}
\horline
\noalign{\vskip2pt}
Burst a & Burst b && $\bar \nu$ & $\% ~>~ 1\sigma$ && $\% ~>~ 2\sigma$
&& $\% ~>~ 3\sigma$ & \cr
\noalign{\vskip2pt}
\horline
\noalign{\vskip2pt}
A at 32\% & B at 32\% && 4.5 & 100.0 && 100.0 && 100.0 & \cr
A at ~8\% & B at 16\% && 2.2 & 90.0 && 84.4 && 78.8 & \cr
A at ~4\% & B at ~8\% && 1.2 & 57.3 & (42.0) &
     27.0 & (19.8) & 11.6 & (8.5) \cr
A at ~2\% & B at ~8\% && 1.1 & 63.2 & (12.2) &
     28.5 & (5.5) & 9.8 & (1.9) \cr
\noalign{\vskip2pt}
\horline
\noalign{\vskip2pt}
A at 32\% & A at 32\% && 16.0 & 0.1 && 0.0 && 0.0 & \cr
A at ~8\% & A at ~8\% && 7.9 & 4.0 && 0.1 && 0.0 & \cr
A at ~4\% & A at ~4\% && 2.5 & 28.5 & (28.0) & 3.9
     & (3.8) & 0.4 & \cr
A at ~2\% & A at ~2\% && 1.1 & 51.3 & (6.0) & 12.0
     & (1.4) & 2.6 & (0.3) \cr
\noalign{\vskip2pt}
\horline
\noalign{\vskip2pt}
B at 32\% & B at 32\% && 3.4 & 16.7 && 0.6 && 0.0 & \cr
B at 16\% & B at 16\% && 1.7 & 28.4 && 2.2 && 0.2 & \cr
B at ~8\% & B at ~8\% && 1.0 & 46.0 & (36.0) & 3.7
     & (2.9) & 0.0 & \cr
\noalign{\vskip2pt}
\horline
}}}}
\bigskip
\baselineskip=12pt
Table 2: Percentage of simulations above a given confidence limit,
based on 1000 simulations of the PSD statistic. The 1$~\sigma$,
2$~\sigma$, and 3$~\sigma$ thresholds correspond to the 68\%, 95.5\%,
and 99.7\% levels for a reduced $\chi^2$ distribution with $\bar \nu$
degrees of freedom.  Percentages are based upon those simulations with
well defined statistics ($\le 1000$). Numbers in parentheses
are based on all 1000 simulations.

\baselineskip=14pt

\vfill\eject

\centerline{\bf Table 3:  Statistical Comparison of Burst Light
Curves}
\bigskip

\def\hrline{\multispan{13}\hrulefill \cr}

\centerline{\vbox{
{\offinterlineskip
\halign{\vrule #  & \strut
  \ \ \hfill # \  & \ \ \hfill # \ & \ \hfill # \hfill \
& \ \ \hfill $#$ \hfill \ & \ \ \hfill $#$ \hfill \ & # \quad
& \ \ \hfill $#$  \ & \ \ \hfill $#$  & # \quad
& \ \ \hfill $#$  \ & \ \ \hfill $#$  \ &\vrule #\cr
\hrline
\noalign{\vskip3pt}
\hrline
\noalign{\vskip2pt}
\omit&
A \quad  &  B \quad  & $\Delta t$ (days)
&  N_f  & \nu &
& (\chi^2_\nu/\nu)_{PSD}  & Q_{PSD} \ &
& (\chi^2_\nu/\nu)_{CPD}  & Q_{CPD}  \ &\omit \cr
\noalign{\vskip2pt}
\hrline
\noalign{\vskip2pt}
\omit&
111$^b$ & 114$^b$ & 0.5 & 2048  &  10 && 10.3 \ \ \ \ &1. \times 10^{-
17} \ &&37.3  \ \ \ \  & 0.00~~~~~ \ &\omit \cr
\omit&
133$^b$ & 1152$^b$ & 221. & 4096 & 10 && 60.5 \ \ \ \ & 0.00~~~~~ \ &&
82.9 \ \ \ \  & 0.00~~~~~ \ &\omit \cr
\omit&
138$^a$ & 444$^a$ & 54.8 & - & - && - \ \ \ \ & -~~~~\ && - \ \ \ \ &
-~~~~\ &\omit \cr %
\omit&
---~ & 575~ & 83.6 & - & - && - \ \ \ \ & -~~~~\ && - \ \ \ \ & -~~~~\
&\omit \cr
\omit&
179~ & 555~ & 69.4 & 256  & 14  && 5.8 \ \ \ \ & 2.\times 10^{- 11} \
&& 4.1  \ \ \ \  &4. \times 10^{-7~ } \ &\omit \cr
\omit&
223~ & 678~ & 82.8 &1024 &4 && 4.3 \ \ \ \ &2. \times 10^{-3~} \ &&5.3
\ \ \ \  &3. \times 10^{-4~} \ &\omit \cr
\omit&
---~ & 824~ & 126. &1024&1&&18.6\ \ \ \ &2.\times 10^{-5~} \ &&16.9 \ \
\ \  &4.\times 10^{-5~} \ &\omit \cr
\omit&
---~ & 1456$^b$ & 283. &1024&3&&4.6\ \ \ \ &3.\times 10^{-3~} \ &&22.3
\ \ \ \  &2.\times 10^{-14} \ &\omit \cr
\omit&
235$^b$ & 1447~ & 280. & 512 & 2 && 6.9 \ \ \ \ & 1. \times 10^{-3~} \
&& 0.9 \ \ \ \ &0.43~~~~~ \ &\omit \cr
\omit&
249~ & 469~ & 28.5 &1024 &21 &&137.\ \ \ \ & 0.00~~~~~ \ &&190.\ \ \ \
& 0.00~~~~~ \ &\omit \cr
\omit&
404~ & 1318~ & 221. & 4096 & 9 && 50.0 \ \ \ \ & 0.00~~~~~ \ &&108.  \
\ \ \  & 0.00~~~~~ \ &\omit \cr
\omit&
444$^a$ & 568$^a$ & 26.2 & - & - && - \ \ \ \ & -~~~~ \ && - \ \ \ \ &
-~~~~ \ &\omit \cr
\omit&
480$^a$ & 1308~ & 202. & - & - && - \ \ \ \ & -~~~~ \ && - \ \ \ \
& -~~~~ \ &\omit \cr
\omit&
493~ & 1430~ & 231. & 128 & 2 && 11.9 \ \ \ \ & 6. \times 10^{-6~} \ &&
0.1 \ \ \ \ &0.89~~~~~ \ &\omit \cr
\omit&
503~ & 676~ & 36.0 & 2048 & 5 && 786.\ \ \ \ & 0.00~~~~~ \ && 36.7 \ \
\ \  & 1.\times 10^{-37} \ &\omit \cr
\omit&
547~ & 1128$^a$ & 135. & - & - && - \ \ \ \ &-~~~~~ \ && - \ \ \ \
& -~~~~~ \ &\omit \cr
\omit&
549~ &398$^b$  & 28.2 & 512 & 5 && 7.6 \ \ \ \ & 4. \times 10^{-7~} \
&& 13.2 \ \ \ \ &8. \times 10^{-13} \ &\omit \cr
\omit&
---~ & 741$^b$ & 42.5 & 512 & 3 && 8.9 \ \ \ \ & 7. \times 10^{-6~} \
&& 6.4 \ \ \ \  &3. \times 10^{-4~} \ &\omit \cr
\omit&
559~ & 1384$^b$ & 204. & 1024 & 3 &&9.5 \ \ \ \ & 3.\times 10^{-6~} \
&&18.3 \ \ \ \  &7. \times 10^{-12} \ &\omit \cr
\omit&
606~ & 1167~ & 133. & 512 & 5 && 6.3 \ \ \ \ & 7. \times 10^{-6~} \ &&
6.2 \ \  \ \ &9. \times 10^{-6} \ &\omit \cr
\omit&
809$^a$ & 1461~ & 165. & - & - && - \ \ \ \ & -~~~~~ \ && - \ \ \ \
& -~~~~~ \ &\omit \cr
\omit&
906~ & 1128$^a$ & 44.2 & - & - && - \ \ \ \ & -~~~~~ \ && - \ \ \ \
& -~~~~~ \ &\omit \cr
\omit&
942~ & 1298~ & 79.7 & 128 & -1 && - \ \ \ \  & -~~~~~ \ && - \ \ \ \
& -~~~~~ \ &\omit \cr
\omit&
1086~ & 1087~ & 0.2 & 512 & 3 && 15.4 \ \ \ \ & 5. \times 10^{-10}\ &&
54.4 \ \  \ \ & 4. \times 10^{-35}\ &\omit \cr
\omit&
1125~ & 1303~ & 49.8 & 512 & 4 && 26.6 \ \ \ \ &3. \times 10^{-17} \
&&16.2 \ \ \ \  & 2. \times 10^{-10} \ &\omit \cr
\omit&
1126~ & 1452~ & 93.3 & 512 & 4 && 25.7 \ \ \ \ & 2. \times 10^{-21} \
&&6.2 \ \ \ \  & 5. \times 10^{-5~} \ &\omit \cr
\omit&
1167~ & 1200~ & 8.1 & 512 & 1 && 21.6 \ \ \ \ & 3.\times 10^{-6~}\ &&
1.4 \ \  \ \ &0.24~~~~~ \ &\omit \cr
\noalign{\smallskip}
\hrline
}}}}

\baselineskip=10pt
\smallskip
\noindent{$^a$ Burst has less than $4$ time bins $3 ~\sigma$ above
noise.}

\noindent{$^b$ Beginning of burst may be missing from data set.}

\baselineskip=12pt

\bigskip

Table 3:  Summary of statistical comparison between burst light
curves.  Bursts were chosen to lie within each others positional
error boxes, as well as have similar durations (\ie the
$t_{50}$ and $t_{90}$ times [\cf Fishman et al. 1992] of one burst
were within a factor of $2$ of those for the other).  Columns A and B
identify bursts by their BATSE trigger
number.  A long dash indicates an entry identical to the preceding
one. $\Delta t$ is the time delay (in days) between bursts. $N_f$ is
the number of Fourier frequency channels used in the comparison. $\nu$
is the degrees of freedom (defined as the number of frequency channels
above noise minus $2$). $(\chi^2_\nu/\nu)_{PSD}$ and
$(\chi^2_\nu/\nu)_{CPD}$ are the reduced $\chi^2$'s derived from the
power spectral density and cross power spectral density, respectively.
$Q_{PSD}$ and $Q_{CPD}$ are the associated probabilities that two
statistically identical signals would show reduced $\chi^2$ as large
or larger than those derived (assuming they obey a
reduced $\chi^2$ distribution).

\bye